\documentclass[conference]{IEEEtran}

\IEEEoverridecommandlockouts

\usepackage{graphicx}
\usepackage{cite}
\usepackage{url}
\usepackage[cmex10]{amsmath}
\usepackage{amssymb}
\usepackage{algorithm}
\usepackage[noend]{algorithmic}
\usepackage{cases}
\usepackage[caption=false,font=footnotesize]{subfig}
\usepackage{color}

\begin{document}

\title{Eigenvalue-based Cyclostationary Spectrum Sensing Using Multiple Antennas}
\author{\IEEEauthorblockN{Paulo Urriza, Eric Rebeiz, Danijela Cabric}\IEEEauthorblockA
{University of California Los Angeles, CA, USA\\
Email: {\{pmurriza, rebeiz, danijela\}}@ee.ucla.edu}
\thanks{This work has been accepted for publication in the proceedings of IEEE GLOBECOM 2012. Copyright may be transferred without notice, after which this version may no longer be accessible.}
}

\maketitle


\begin{abstract}
In this paper, we propose a signal-selective spectrum sensing method for cognitive radio networks and specifically targeted for receivers with multiple-antenna capability. This method is used for detecting the presence or absence of primary users based on the eigenvalues of the cyclic covariance matrix of received signals. In particular, the cyclic correlation significance test is used to detect a specific signal-of-interest by exploiting knowledge of its cyclic frequencies. The analytical threshold for achieving constant false alarm rate using this detection method is presented, verified through simulations, and shown to be independent of both the number of samples used and the noise variance, effectively eliminating the dependence on accurate noise estimation. The proposed method is also shown, through numerical simulations, to outperform existing multiple-antenna cyclostationary-based spectrum sensing algorithms under a quasi-static Rayleigh fading channel, in both spatially correlated and uncorrelated noise environments. The algorithm also has significantly lower computational complexity than these other approaches.
\end{abstract}



\section{Introduction}
\label{sec:Introduction}

Spectrum sensing is a key step in effectively realizing cognitive radio networks (CRN). In the CR access paradigm, secondary users (SU) in a CRN are allowed to access spectrum reserved for use by licensed or primary users (PU) given that 1) those resources are either currently unoccupied or 2) interference to the primary network is kept under an acceptable level \cite{Haykin2005}. The main goal of spectrum sensing is to accurately and efficiently detect the presence or absence of a PU in a given band.

Several spectrum sensing methods have been proposed in the literature \cite{Yucek2009}. In general, these methods can be categorized as being based on either energy detection, spectral correlation (cyclostationarity), or matched filtering. Energy detection requires the least prior knowledge about the signal, while matched filtering requires the most. Spectral correlation-based techniques lie in between, requiring either prior knowledge or accurate estimation of the cyclic frequencies present in the PU transmission signal. Although energy detection offers the lowest computational complexity and is the optimal blind detector in the presence of i.i.d. noise, its performance relies on accurate knowledge of noise power due to the SNR wall phenomenon \cite{Tandra2008}. The detection performance of energy detection also degrades in a temporally correlated noise environment.

In some scenarios, spectral correlation-based methods offer several advantages over other spectrum sensing approaches. Unlike energy detection, they do not suffer from the SNR wall issue. These methods are also resilient to temporally correlated noise and enable signal-selective spectrum sensing where the presence a signals-of-interest (SOI) can be detected based on their unique cyclic features due to their modulation type, symbol rate, and carrier frequency \cite{Gardner1987a}. 

One issue encountered with all spectrum sensing methods is the effect of a fading in the channel between the PU and SU. There is some probability of low detection performance whenever the channel is in a deep fade. This can be alleviated by exploiting spatial diversity either through the use of cooperative spectrum sensing \cite{Quan2008} or if available, the use of multiple antennas. As a result, spectrum sensing algorithms exploiting multiple antennas have received considerable interest \cite{Taherpour2010,Tugnait2012}. 

Algorithms that leverage the cyclostationarity property have been applied in the past for multiple antenna receivers. In \cite{Sadeghi2008}, the sum of the individual cyclic correlation for each antenna was proposed. Such methods are considered \textit{post-combining} techniques since knowledge of the channel state information (CSI) is not exploited. On the other hand, \textit{pre-combining} techniques which utilize an estimate of the CSI to varying degrees have been shown to have better performance. A method based on equal gain combining (EGC), was investigated in \cite{Chen2008}. This method uses phase offset estimates to align the raw samples from each antenna. The aligned signals are then summed before finding the cyclic correlation. Finally, a Blind Maximal Ratio Combining (BMRC) scheme was evaluated in \cite{Jitvanichphaibool2010a} which utilized the singular value decomposition (SVD) to find an estimate of the CSI and applied MRC on the raw samples.

In this paper we propose a spectrum sensing algorithm designed for use in a multiple antenna system based on the cyclic correlation significance test (CCST). The CCST was used in \cite{Schell1990a} to perform cyclostationary source enumeration using an information-theoretic criterion. However, the performance of this statistic in the context of multiple-antenna cyclostationary spectrum sensing has not been investigated in prior work. The performance of this method in fading channels has also not been evaluated. In this paper, we show that the proposed method has better detection performance than existing methods and has less computational complexity than BMRC from \cite{Jitvanichphaibool2010a}. We investigate the algorithm's performance in both uncorrelated and correlated noise environments. The scheme's robustness to intereference is also shown.

The rest of the paper is organized as follows. The system model is introduced in Section \ref{sec:Model} including a brief discussion of cyclostationarity. The proposed algorithm is detailed in Section \ref{sec:Proposed}. Numerical results for various scenarios are presented in Section \ref{sec:Results}. Finally, the paper is concluded in Section \ref{sec:Conclusion}.


\section{Background and System Model}
\label{sec:Model}

\subsection{Background on Cyclostationarity}

A signal is considered to be cyclostationary if its statistical properties are periodic. Equivalently, if the cyclic autocorrelation function, defined as:
\begin{equation}
\label{eqn:CyclicCovariance}
R_{x}^{\alpha}(\tau)=\!\!\!\lim_{\Delta t\rightarrow\infty}\frac{1}{\Delta t}\!\int_{-\frac{\Delta t}{2}}^{\frac{\Delta t}{2}}\!\!x\!\left(t+\frac{\tau}{2}\right)\!x^{*}\!\left(t-\frac{\tau}{2}\right)\!e^{-j2\pi\alpha t}dt,
\end{equation}
is non-zero with some $\tau$ for at least one $\alpha \neq 0$, the signal is said to exhibit second-order cyclostationary property with $\alpha$ referred to as the cyclic frequency.

For example, in BPSK signals, cyclostationary features exist at $\alpha=\frac{k}{T_b}$ and at $\alpha=\pm 2 f_c + \frac{k}{T_b}$, where $T_b$ is the symbol period, $f_c$ is the carrier frequency, and $k\in\{-1,0,1\}$. Detailed analysis of the cyclostationary features for various digital modulations can be found in \cite{Gardner1987a}.

\subsection{Signal Model and Assumptions}
We adopt a similar signal model as that used in \cite{Jitvanichphaibool2010a}. The spectrum sensing problem is to decide between two hypotheses: $\mathcal{H}_0$, where the signal is absent; and $\mathcal{H}_1$, where it is present. The received signal samples under the two hypothesis are given respectively as follows:
\begin{equation}
x(n)=\begin{cases}
\eta(n), & \mathcal{H}_0\\
s(n) + \eta(n), & \mathcal{H}_1.
\end{cases}
\end{equation}

The received signal, sampled at a rate of $1/T_s$, forms $M$ streams coming from each antenna with $N$ samples each. This received signal is defined as $
\mathbf{x}(n)\triangleq[x_{1}(n),x_{2}(n),\ldots,x_{M}(n)]^{T}$. The received signal is the superposition of $P$ signal sources (including both the SOI and any interferers) and can be expressed in vector form as
\begin{equation}
\label{eqn:ReceivedDecompose}
\mathbf{x}\left(n\right)=\sum_{j=1}^{P}\mathbf{h}_{j}\left(n\right)\otimes s_{j}\left(n\right)+\boldsymbol{\eta}\left(n\right),
\end{equation}
where $\otimes$ is the convolution operation over $n$ and $\boldsymbol{\eta}(n)$ is the receiver noise denoted by $\boldsymbol{\eta}(n)
\triangleq[\eta_{1}(n),\eta_{2}(n),\ldots,\eta_{M}(n)]^{T}$, where every $\eta_i$ is a purely stationary Gaussian random process ($R_{\eta}^{\alpha}(\tau)=0$ for any $\alpha\neq 0$) with variance of $\sigma^2_\eta$. For simplicity, we restrict that only one PU transmission, $s_1(n)$, is considered a SOI and that it is cyclostationary with a unique cyclic frequency at $\alpha=\alpha_0$. The other sources are considered interferers. The channel experienced by each of the $P$ sources is given by $\mathbf{h}_{j}(n)\triangleq[h_{j1}(n),h_{j2}(n),\ldots,h_{jM}(n)]^{T}$, where $h_{jk}(n)$ is the channel between the $j$th source and the $k$th antenna. We assume that the channel, although unknown to the receiver, stays constant over the frame of observation.

\subsection{Spatially Correlated Noise Environments}

In the case of spatially correlated noise, which can happen when there is substantial ambient noise in the band, we model $\boldsymbol{\eta}(n)$ to have a covariance matrix  given by $\{\mathbf{R}_{\boldsymbol{\eta\eta}}\}_{ij}=\sigma_{\eta}\rho^{\left|i-j\right|}$. Thus with $\rho=0$, the covariance matrix simplifies to an identity matrix giving spatially white noise, while $\rho=1$ gives fully correlated noise over all antennas. Varying degrees of partial correlation can be achieved by setting $0<\rho<1$.


\section{Proposed Algorithm}
\label{sec:Proposed}
In this section, we describe the proposed method. We focus on a single cycle frequency detection, but this approach could be generalized to multi-cycle detection. The key idea of this detection algorithm is based on the theory of canonical variates or theory of common factors. As discussed in \cite{Lawley1959}, and subsequently utilized in \cite{Schell1990a},  the number of common factors between two $M\times1$ time-series vectors $\mathbf{u}(n)$ and $\mathbf{v}(n)$ can be inferred from the rank of the matrix
\begin{equation}
\label{eqn:CommonFactors}
\mathbf{R}=\mathbf{R}_{\mathbf{vv}}^{-1}\mathbf{R}_{\mathbf{vu}}\mathbf{R}_{\mathbf{uu}}^{-1}\mathbf{R}_{\mathbf{uv}},
\end{equation}
where the covariance matrices are defined as the time average $\mathbf{R}_{\mathbf{uv}}=\frac{1}{N}\sum_{n=0}^{N-1}\mathbf{u}(n)\mathbf{v}^H(n)$. In the non-asymptotic case, when this matrix is always full rank, a threshold on the eigenvalues can be applied to determine the asymptotic rank with a given confidence level. The same criterion, referred to as the Cyclic Correlation Significance Test (CCST), was applied for cyclostationary source enumeration in \cite{Schell1990a} by taking $\mathbf{u}(n)=\mathbf{x}(n)$ and $\mathbf{v}(n)=\mathbf{x}(n-\tau)e^{-j2\pi\alpha nT_s}$ for a given lag $\tau$ and cyclic frequency $\alpha$. Additionally, some cyclic frequencies, such as those located on $\alpha=\pm 2f_c$ for BPSK, only appear in the conjugate cyclic correlation matrix by instead taking $\mathbf{v}(n)=\mathbf{x}^*(n-\tau)e^{-j2\pi\alpha nT_s}$.

We further adapt the CCST into the binary hypothesis test required for signal-selective spectrum sensing. Under $\mathcal{H}_0$, (\ref{eqn:CommonFactors}) will have zero rank as $N\rightarrow\infty$. Thus, by applying a threshold to all the eigenvalues we can infer the presence or absence of the PU of interest using a $M$ antenna receiver. Prior to performing the detection, we pick the lag $\tau$ that provides the best detection performance based on the modulation format used by the PU. This could be done off-line by performing the maximization, $\tau_0=\arg\max_\tau |R_s^{\alpha_0}(\tau)|$.

The steps of the algorithm are summarized as follows:
\begin{enumerate}
	\item Estimate the covariance matrix of size $M\times M$
	\begin{equation}
	\label{eqn:Covariance}
	\hat{\mathbf{R}}_{\mathbf{xx}}(\tau_0)=\frac{1}{N}\sum_{n=0}^{N-1-\tau_0}\mathbf{x}\left(n\right)\mathbf{x}^{H}\left(n-\tau_0\right).
	\end{equation}
	\item Estimate the cyclic correlation matrix using a cyclic cross-correlogram at cyclic frequency $\alpha$ and lag $\tau_0$, defined as
	\begin{equation}
	\label{eqn:CyclicCovariance}
	\hat{\mathbf{R}}_{\mathbf{xx}}^{\alpha_0}(\tau_0)=\frac{1}{N}\!\!\sum_{n=0}^{N-1-\tau_0}\!\!\!\!\!\mathbf{x}\left(n\right)\mathbf{x}^{H}\!\!\left(n-\tau_0\right)e^{-j2\pi\alpha_0 nT_s}.
	\end{equation}
	We will refer to the $\tau_0$-lag covariance matrices for both conventional and cyclic autocorrelation function simply as $\hat{\mathbf{R}}_{\mathbf{xx}}$ and $\hat{\mathbf{R}}_{\mathbf{xx}}^{\alpha_0}$ from this point for the sake of brevity, since other $\tau$ are not utilized by the proposed algorithm. The dependence on $\tau$ will be indicated explicitly whenever necessary. The CCST is then calculated by finding the matrix
	\begin{equation}
	\label{eqn:Metric}
	\hat{\mathbf{R}}=\hat{\mathbf{R}}_{\mathbf{xx}}^{-1}\hat{\mathbf{R}}_{\mathbf{xx}}^{\alpha_0} \hat{\mathbf{R}}_{\mathbf{xx}}^{-1}\hat{\mathbf{R}}_{\mathbf{xx}}^{\alpha_0 H}.
	\end{equation}
	\item Calculate its singular value decomposition (SVD)
	\begin{equation}
	\label{eqn:SVD}
		\hat{\mathbf{R}} = \mathbf{U \Sigma V}^H,
	\end{equation}
	where $\mathbf{U}$ and $\mathbf{V}$ are unitary matrices whose columns are the left and right singular vectors, respectively and $\mathbf{\Sigma}=\mathbf{I}_M\left[\mu_1,\mu_2,\ldots,\mu_M\right]^{T}$ is the diagonal matrix of real, non-negative singular values, $\mu_i$, and $\mathbf{I}_M$ is the identity matrix of size $M \times M$.
	\item Compute the test statistic by taking
	\begin{equation}
	\label{eqn:TestStatistic}
	\mathcal{T^{\alpha}_{\mathbf{xx}}}=-N\ln\prod_{i=1}^{M}\left(1-\mu_{i}\right).
	\end{equation}
	The factor $N$ is used to scale the test statistic so that its distribution is independent of the number of samples used.
	\item Decision: 
	$\mathcal{T^{\alpha}_{\mathbf{xx}}}\mathop{\gtrless}_{\mathcal{H}_0}^{\mathcal{H}_1}\gamma$,
	where $\gamma>0$ is a threshold chosen to achieve constant false alarm rate (CFAR) which will be discussed in the following section.
	
\end{enumerate}

Note that all $\mathbf{x}^H(n)$ can be replaced with $\mathbf{x}^T(n)$ if the conjugate cyclic correlation matrix is needed. We refer to these two metrics as the non-conjugate cyclic correlation significance test (NC-CCST) and the conjugate cyclic correlation significance test (C-CCST) respectively.

\subsection{Probability of False Alarm and Threshold Computation}

Two key parameters are used to evaluate the performance of spectrum sensing algorithms. The detection probability or $P_d$ is the probability of being at $\mathcal{H}_1$ and accurately detecting the PU ($P_d=\Pr(\mathcal{T^{\alpha}_{\mathbf{xx}}}>\gamma \mid \mathcal{H}_1)$). On the other hand, the false alarm probability, $P_{fa}$, is the probability of being at $\mathcal{H}_0$ and mistakenly detecting a PU ($P_{fa}=\Pr(\mathcal{T^{\alpha}_{\mathbf{xx}}}>\gamma \mid \mathcal{H}_0)$).

It has been shown in \cite{Lawley1959} that the limiting distribution ($N\rightarrow\infty$) of the test statistic based on the NC-CCST in (\ref{eqn:TestStatistic}) approaches a $\chi^2$ distribution with degree-of-freedom $M^2$. Following a similar proof, it can be shown that the distribution for the C-CCST is also $\chi^2$ but with degree-of-freedom $M(M+1)$.

Based on these asymptotic distributions of $\mathcal{T^{\alpha}_{\mathbf{xx}}}$ under $\mathcal{H}_0$, the detection threshold $\gamma$ can be set to achieve a desired $P_{fa}$ by satisfying
\begin{equation}
\label{eqn:threshold}
\int_{\gamma}^{\infty}f_{\chi_{k}^{2}}\left(x\right)dx=P_{fa},
\end{equation}
where
\begin{equation}
\\k=\begin{cases}
M^{2} & \text{NC-CCST}\\
M(M+1) & \text{C-CCST}
\end{cases},
\end{equation}
and $f_{\chi_{k}^{2}}(\cdot)$ is the probability density function (pdf) of a $\chi^2$ random variable with degree-of-freedom $k$.

\begin{figure}
\centering
\includegraphics[width=\columnwidth]{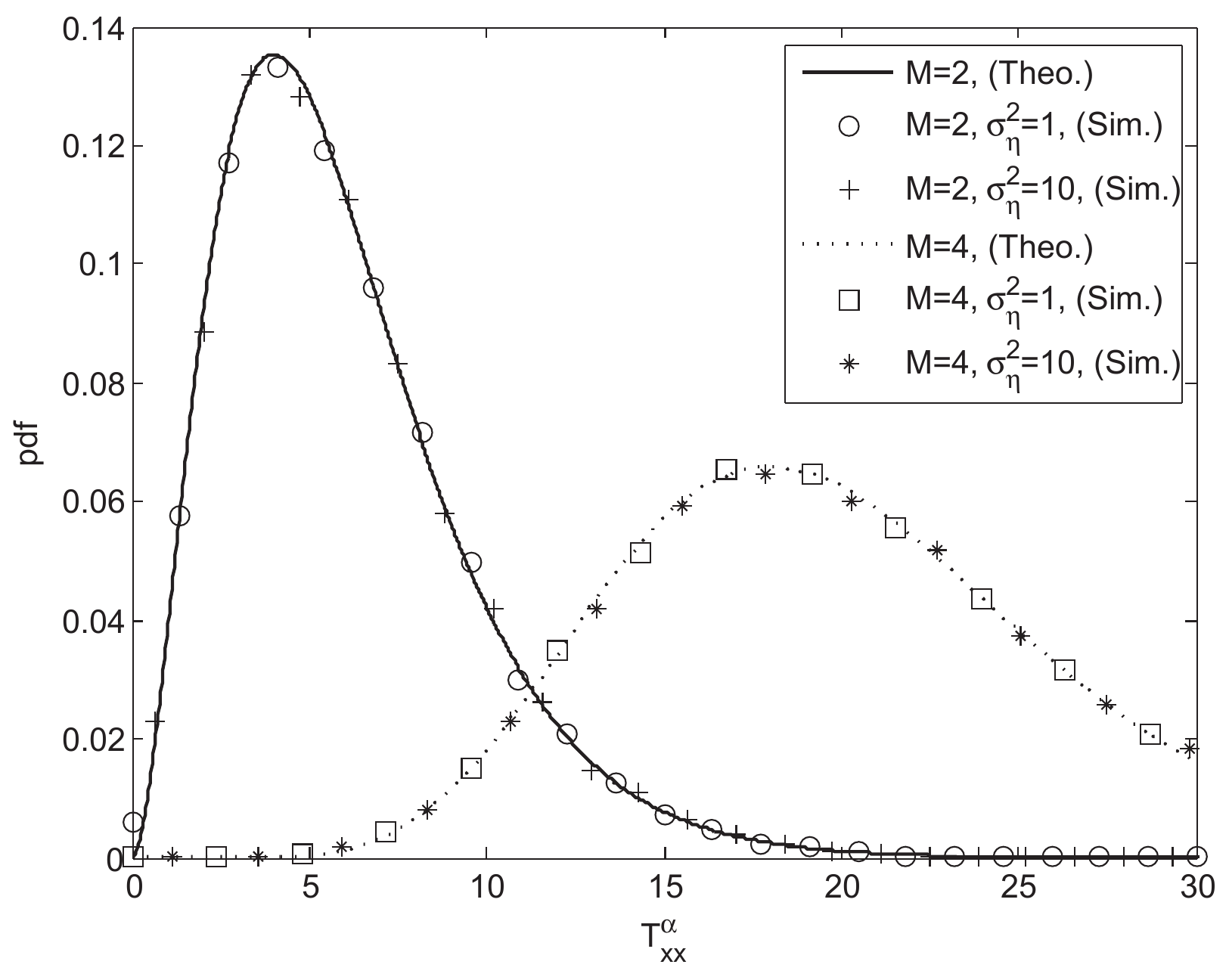}
\caption{Verification of the asymptotic distribution of the proposed test statistic (C-CCST) under $\mathcal{H}_0$ with different number of antennas ($M=\{2,4\}$). Simulated pdfs for different noise variances ($\sigma^2_\eta=\{1,10\}$) show both the accuracy of the approximation and the test statistic's independence to noise power. The same number of samples $N=4000$ per antenna is used.}
\label{fig:pdf}
\end{figure}

These asymptotic distributions are verified to closely match simulation in Fig.~\ref{fig:pdf} for $N=4000$. Due to the scaling factor in (\ref{eqn:TestStatistic}), the distribution is independent of $N$ if the number of samples is large enough. The empirical pdfs for two different $\sigma^2_\eta$ values are also shown to demonstrate how the test statistic's distribution under $\mathcal{H}_0$ is independent of noise power.

\subsection{Comparison With Existing Approaches}
\label{subsec:others}
The algorithms for multiple-antenna spectrum sensing that exploit cyclostationarity that are currently in the literature can generally be classified into two categories. Techniques that do not exploit any knowledge of the CSI are referred to as \textit{post-combining} methods. The simplest method to do this is to find the sum of the spectral correlation test statistic estimated individually from each antenna \cite{Sadeghi2008}. We refer to this approach as SUM-MSDF (where MSDF means Modified Spectral Density Function). The MSDF is defined as the spectral correlation function normalized by signal energy as discussed in \cite{Jitvanichphaibool2010a}.

Another existing approach is to sum the raw signals from each antenna and then perform a single spectral correlation test. However, we encounter a problem when the channel is not simply AWGN but instead has random fading. In this case, each antenna will have some unknown phase offset and attenuation. Thus, simply adding the raw signals non-coherently would decrease the probability of detection. This problem is addressed in \cite{Chen2008} by first aligning the phase of each antenna. An estimate of the relative phase difference between each antenna is calculated by finding both the cyclic correlation of one antenna chosen as reference (auto-SCF) and the cross-cylic correlation of every other antenna and the reference antenna. The phase difference can then be extracted from these two. We refer to this method in our comparisons as EGC.

Finally, MRC is used in \cite{Jitvanichphaibool2010a}. Blind channel estimation is achieved by taking the vector corresponding to the highest singular value of (\ref{eqn:CyclicCovariance}) as an estimate of the channel, $\hat{\mathbf{h}}$. The raw samples from each antenna are combined using
\begin{equation}
	y(n) = \frac{\hat{\mathbf{h}}^H\mathbf{x}(n)}{\Vert \hat{\mathbf{h}}\Vert } .
\end{equation}
The cyclic correlation test is then performed on the combined samples $y(n)$. This method is called MSDF with Blind Maximal Ratio Combining or BMRC-MSDF. It was shown to outperform the other techniques but at the cost of additional complexity due to the channel estimation and combining. One issue with this approach is the fact that the cyclic correlation is calculated twice. The first is used to blindly estimate the channel and the second to perform the detection on the combined samples. In contrast, the method proposed in this paper, which we refer to as eigenvalue-based cyclostationary spectrum sensing or EV-CSS only needs to perform the first part of BMRC-MSDF, the SVD, and uses the singular values themselves to infer the presence or absence of the PU.
 
\subsection{Advantages of the Proposed Algorithm}

As with other cyclostationarity-based spectrum sensing algorithms, one major advantage of the proposed method is its robustness to the noise uncertainty problem. Since the noise is assumed to be stationary and does not exhibit cyclostationarity at any $\alpha \neq 0$, its cyclic correlation approaches zero as $N\rightarrow\infty$. Thus, the effect of any error in the noise power estimate on the detection probability can be eliminated by taking more samples. However, in the interest of conserving power and arriving at a timely decision, both of which are high priority in the case of CR applications, we aim to minimize $N$ needed to achieve a target $P_d$. This presents another, more subtle, issue related to noise uncertainty. 

In the non-asymptotic scenario, the SCF under $\mathcal{H}_0$ has been shown to depend on both $N$ and the noise power $\sigma_\eta$ \cite{Rebeiz2011}. Therefore, the proper detection threshold is still a function of the noise variance. By incorrectly specifying this threshold, the detector could be at the wrong point in the receiver operating characteristic (ROC) curve. Equivalently, the target CFAR cannot be achieved. However, as previously discussed and demonstrated in Fig.~\ref{fig:pdf}, the proposed test statistic is independent of both $\sigma^2_\eta$ and $N$. Consequently, the threshold $\gamma$ only needs to be chosen once for a given number of antennas $M$ to guarantee CFAR. This property has been shown for other eigenvalue-based approaches \cite{Zeng2009}. It derives from the fact that noise power estimation is built-in to the algorithm.

\subsection{A Note on Complexity}
\label{subsec:complexity}
We can make an approximate complexity comparison of the proposed algorithm with the best performing existing algorithm (BMRC-MSDF) by taking number of complex multiplications required for each under the same number of samples $N$. Since the cyclic covariance operation and the SVD are common to both algorithms they are not included in the analysis.

Assuming the MSDF is calculated using an $N_S$-point Fast Fourier Transform (FFT) it requires in the order of $N\log_2(N_S)$ multiplications. In addition, $(M+1)N$ multiplications are needed to perform the MRC and normalization. Finally, the correlation in frequency uses $NN_S/2$ multiplications. Thus, the BMRC-MSDF approach performs in the order of $N(\log_2(N_S)+N_S/2+M+1)$ multiplications without the SVD and the cyclic covariance.

In comparison, the proposed EV-CSS method finds the conventional covariance in addition to the same SVD and cyclic covariance as BMRC-MSDF, or in the order of $NM^2$ multiplications. The operation $\hat{\mathbf{R}}_{\mathbf{xx}}^{-1}\hat{\mathbf{R}}_{\mathbf{xx}}^{\alpha}$ in (\ref{eqn:Metric}) is essentially the solution to a generalized linear system which can be seen as an LU decomposition requiring approximately $2M^3/3$ multiplications. Therefore, the EV-CSS approach requires in the order of $NM^2+2M^3/3$ multiplications in addition to the common operations with BMRC-MSDF. Since $M$ is typically much less than both $N$ and $N_S$, there is overall a significant decrease in complexity with the proposed algorithm. For example, if we take $N=4000$, $N_S=128$, and $M=2$, (same parameters used in \cite{Jitvanichphaibool2010a}), the BMRC-MSDF requires $\sim$296K multiplications while EV-CSS needs only $\sim$16K multiplications, without counting the common operations.


\section{Numerical Results and Discussion}
\label{sec:Results}

In this section, simulation results are presented in order to compare the proposed algorithm with the various existing techniques discussed in Section~\ref{subsec:others}. The PU is assumed to be transmitting a BPSK signal at a carrier frequency $f_c=80$ KHz with symbol period of 25 $\mu$s. The SU is assumed to have $M=2$ antennas unless otherwise specified. Each antenna is sampled at a rate $f_s=320$ kHz. For CFAR experiments we set $P_{fa}=0.1$.

In all experiments, the channel between the PU and each antenna of the SU, $\mathbf{h}$, is modeled as a quasi-static Rayleigh fading channel, where the fading remains constant during the whole frame of $N=4000$ samples per antenna used for detection. This can be described using a channel vector for the $i$th frame as
$\mathbf{h}_i=[r_1e^{j\theta_1},r_2e^{j\theta_2},\ldots,r_Me^{j\theta_M}]^{T}$,
where $r_n$ is a Rayleigh distributed random variable of unit variance and $\theta_n$ is a uniformly distributed random variable in $[0,2\pi]$. We ignore dispersive channels where each element of $\mathbf{h}_i$ is modeled as a multi-tap filter, $\mathbf{h}_i(n)$, since cyclostationary spectrum sensing has already been shown to be robust to the effects of temporal correlation in prior work \cite{Jitvanichphaibool2010a}.

For all algorithms, the same cyclic frequency located at $\alpha_0=2f_c$ is used. This cyclic feature has been shown to only be present in the conjugate cyclic autocorrelation which means the C-CCST statistic is used for the proposed algorithm. This feature is chosen because of its relative strength compared to other cyclic frequencies. The maximum cyclic autocorrelation at this cyclic frequency is observed at $\tau_0=0$. As for the algorithm that computes the MSDF, the frequency resolution is chosen to be $f_s/100$.

\subsection{Spatially Uncorrelated Noise}

\begin{figure}
\centering
\includegraphics[width=\columnwidth]{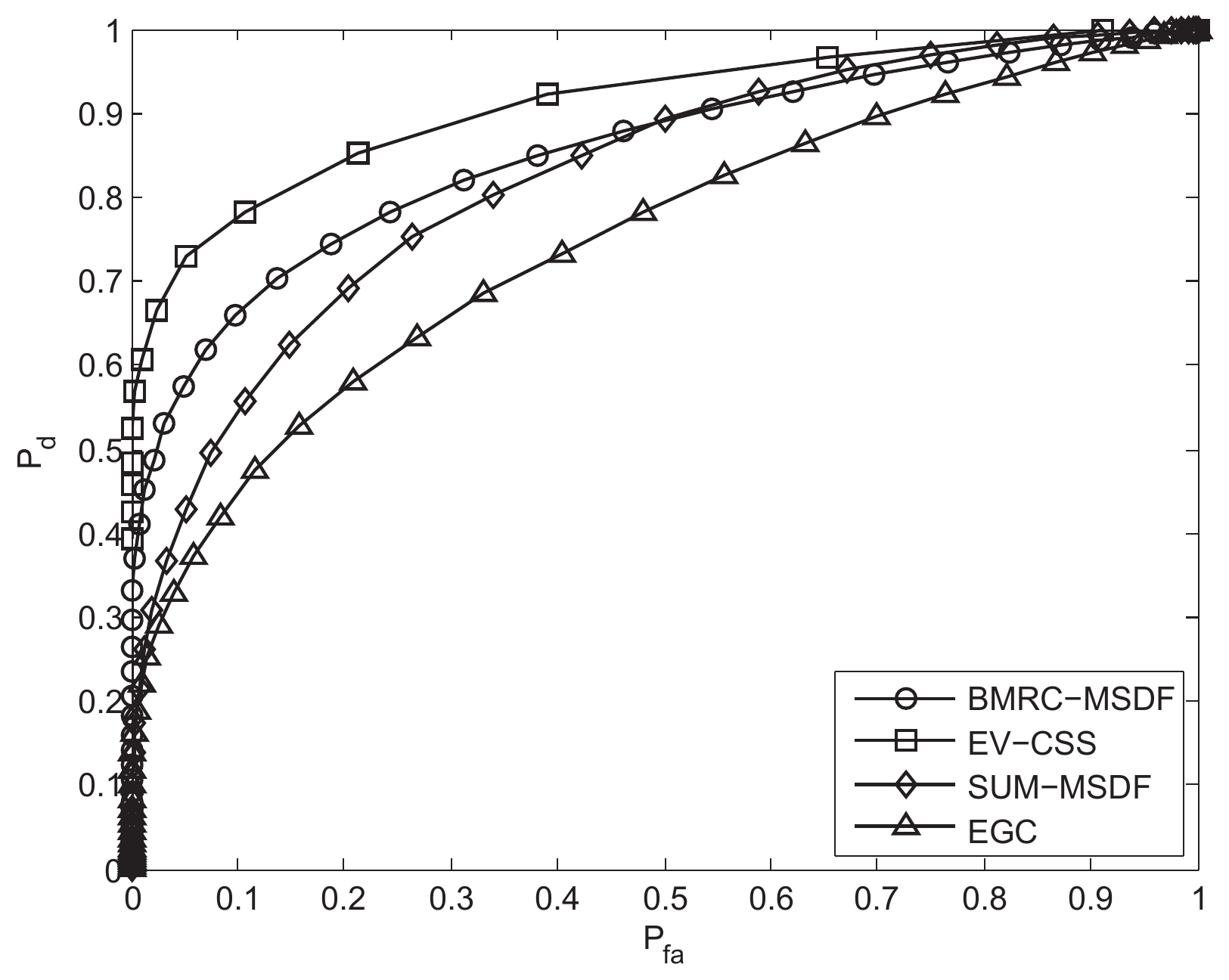}
\caption{ROC curves comparing the proposed EV-CSS to BMRC-MSDF and to a post-combining method using EGC. The same number of samples $N=4000$ and $\text{SNR}=-14$ dB are used for all techniques.}
\label{fig:ROC}
\end{figure}

In the case where the noise between each antenna is uncorrelated ($\mathbf{R}_{\boldsymbol{\eta\eta}}=\sigma_\eta\mathbf{I}_M$) and only the SOI occupies the observed band, the ROC curve completely determines the performance of each detection algorithm under a given SNR and $N$. A comparison of these ROC curves for the algorithms discussed are shown in Fig.~\ref{fig:ROC}. For these simulations the SNR is fixed at -14 dB, although the trends are consistent for other SNR in the range $[-20,0]$ dB where the experiments were conducted. As seen from this figure, the proposed method has better detection performance than any of the other algorithms.

Interestingly, the method also outperforms BMRC-MSDF which as discussed in Section~\ref{subsec:complexity} has significantly higher computational complexity. Although this result initially appears to be counter-intuitive, further experiments show that in the case of perfect CSI (not shown in figure) the MRC-MSDF has comparable performance to EV-CSS. Therefore, we can conclude that at very low SNR the blind channel estimates based on the SVD have large errors and the full benefit of MRC is not achieved. In contrast, the EV-CSS is able to fully take advantage of the information from all antennas because it does not directly use an estimate of the CSI.

\begin{figure}
\centering
\includegraphics[width=\columnwidth]{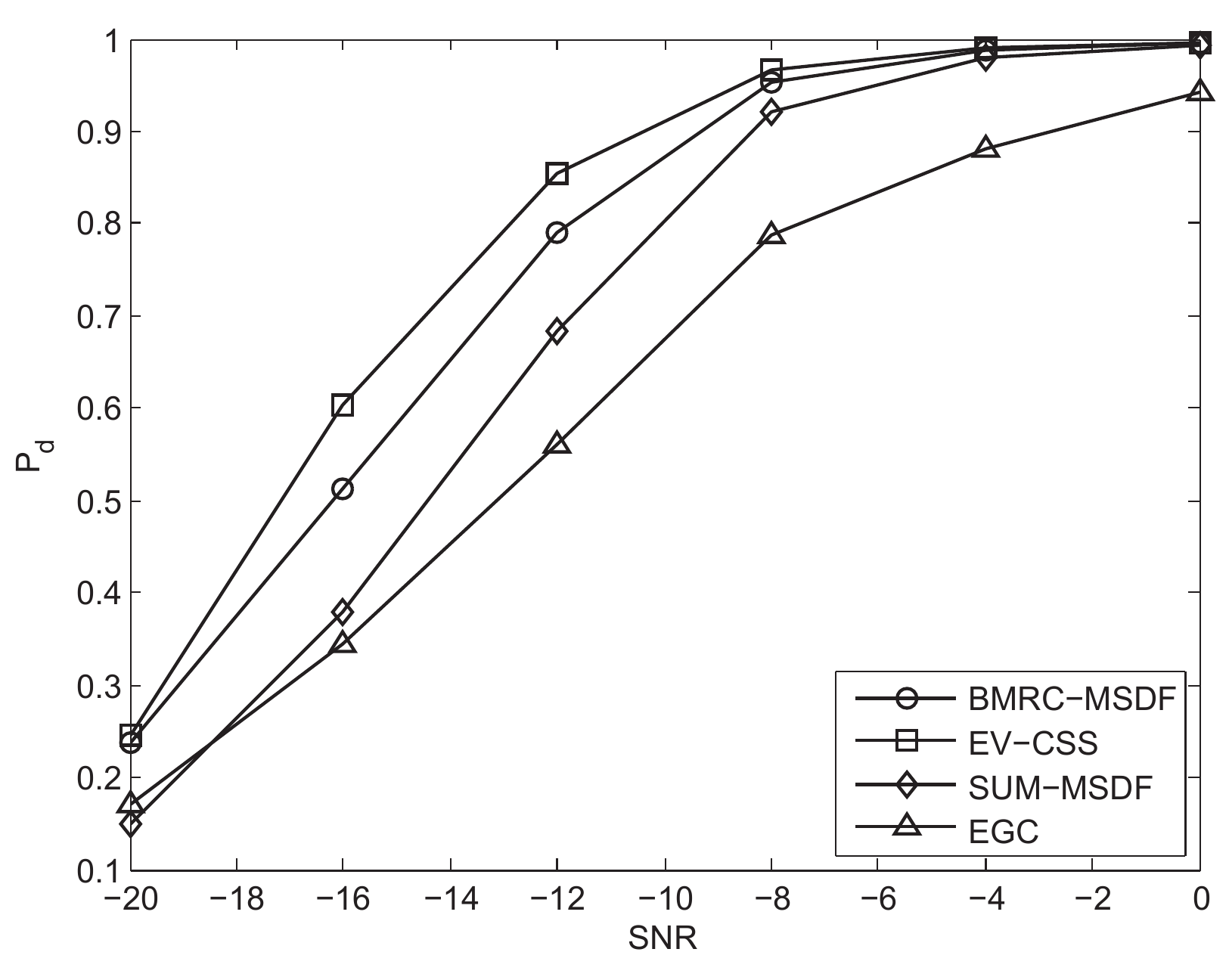}
\caption{The effect of varying SNR to detection probability ($P_d$) for different multiple antenna cyclostationary spectrum sensing techniques including the proposed EV-CSS under ($P_{fa}=0.1$) under uncorrelated noise. The same number of samples $N=4000$ is used.}
\label{fig:pd}
\end{figure}

The effect of varying SNR on probability of detection is shown in Fig.~\ref{fig:pd}. The $P_{fa}$ is kept constant at 0.1. It can be seen that all the methods except EGC reach the $P_d=1$ at 0 dB SNR. Again, we see that the proposed method consistently has better detection performance at all SNRs compared to the rest of the algorithms. 

\subsection{Spatially Correlated Noise}

\begin{figure}
\centering
\includegraphics[width=\columnwidth]{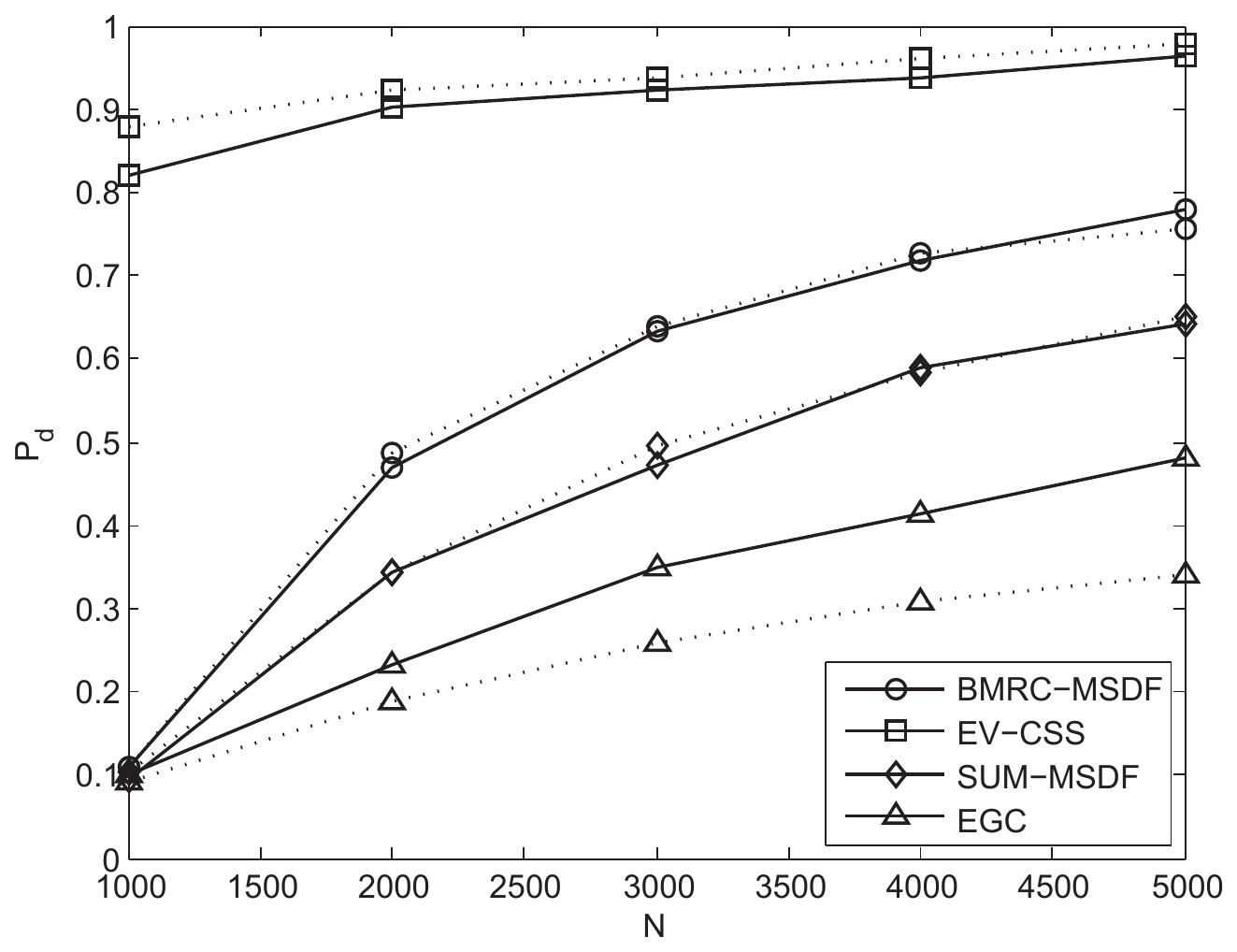}
\caption{The effect of varying number of samples, $N$, to detection probability ($P_D$) of the various techniques. Solid lines are for spatially uncorrelated noise environments while dashed lines are for $\rho=0.5$.}
\label{fig:pdcorr}
\end{figure}

The effect of varying number of samples, $N$, is shown in Fig.~\ref{fig:pdcorr}. The second set of plots also show a spatially correlated noise environment with $\rho=0.5$. All methods, with the exception of EGC, are robust to spatial correlation and in fact, the proposed method shows a slight improvement in spatially correlated noise. We can also see that the proposed method shows good performance even at low number of samples while the MSDF based methods require more samples to achieve comparable performance.

\subsection{Effect of Interfering Signal}

\begin{figure}
\centering
\includegraphics[width=\columnwidth]{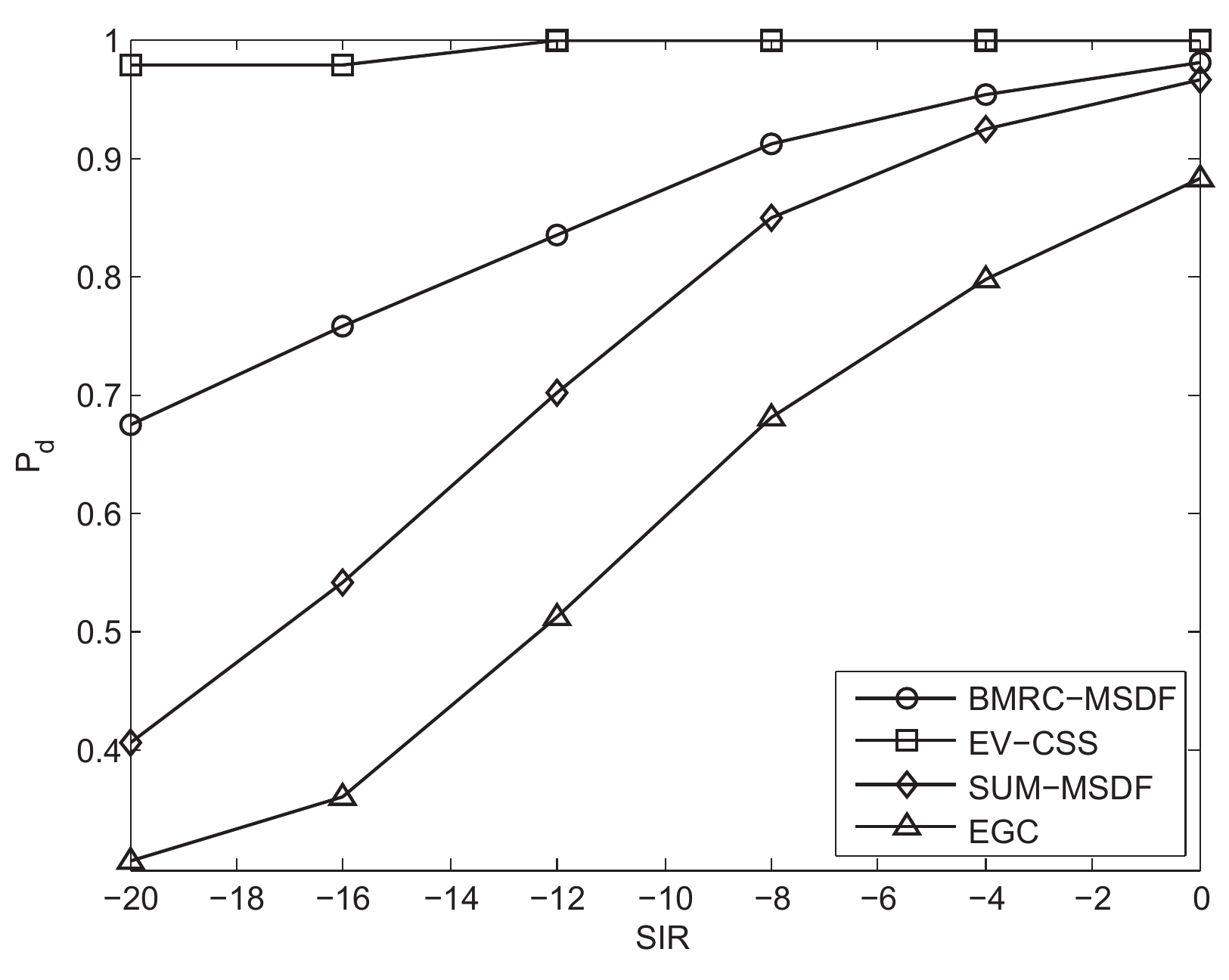}
\caption{The effect a co-channel BPSK interfer on detection probability with 30\% spectral overlap. The number of samples used for all antennas is $N=4000$ and the noise level is kept constant at $\sigma_\eta=1$.}
\label{fig:interfer}
\end{figure}

We test the robustness of these algorithms in the presence of a strong co-channel interferer by introducing another BPSK signal with the same symbol rate with 30\% spectral overlap. The effect of the interferer on the detection performance is shown in Fig.~\ref{fig:interfer} as the signal-to-interferer ratio (SIR), defined as the ratio of the interferer power to SOI power, is varied from -20 dB to 0 dB. The noise is kept constant at $\sigma_\eta=1$. The proposed algorithm shows very good signal selectivity, giving $P_d$ close to 1 even in the presence of a co-channel inteferer with 100 times the SOI's power. By performing the correlation entirely in time domain, the proposed method is able to suppress the interferer much better than the MSDF.

\subsection{Effect of Number of Antennas}

\begin{figure}
\centering
\includegraphics[width=\columnwidth]{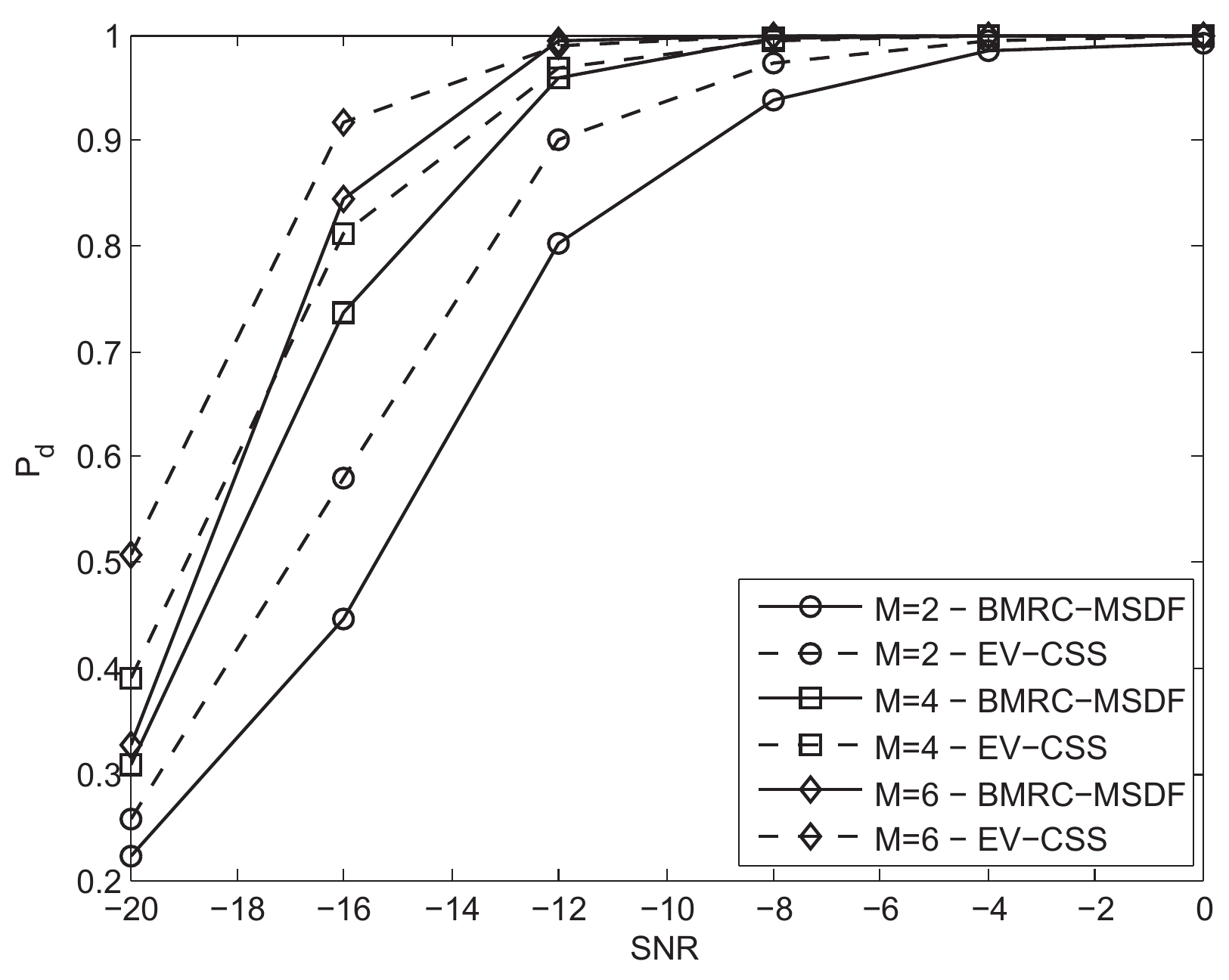}
\caption{The effect of the number of antenna on the detection probability. The same number of samples per antenna $N=4000$ is used.}
\label{fig:varym}
\end{figure}

The effect of number of antennas, $M$, on detection accuracy is studied in Fig.~\ref{fig:varym}. Note that for EV-CSS, to keep the the $P_{fa}$ constant at 0.1 the threshold must be set to a new value based on (\ref{eqn:threshold}). On the other hand, for the rest of the algorithms, the threshold is set for different $N$, $M$, and SNR. The SNR across all antennas is assumed to be the same. Since the total number of samples increases with more antennas ($N_T=MN$), we expect both algorithms to perform better with higher $M$. In addition, more antennas also introduce spatial diversity which reduces the probability of all antennas being at a deep fade during the sensing period. Similar to previous results, the EV-CSS has better performance than BMRC-MSDF for different values of $M$.

\section{Conclusion}
\label{sec:Conclusion}
A multi-antenna cyclostationary-based spectrum sensing algorithm based on the cyclic correlation significance test was proposed. The method was shown to outperform current multiple antenna signal-selective spectrum sensing methods in the literature. The computational complexity of the algorithm was also compared with that of the best performing existing algorithm that uses MRC by blindly estimating the CSI and was shown to require substantially less multiplications. The detection threshold for CFAR was also determined both theoretically and via simulation to be independent of the noise variance or the number of samples. This means that a single threshold is required for a given number of antenna, eliminating the need for separate noise estimation. Future work includes theoretical performance analysis of the proposed algorithm.



\vspace{-3mm}
\begin{thebibliography}{10}
\providecommand{\url}[1]{#1}
\csname url@samestyle\endcsname
\providecommand{\newblock}{\relax}
\providecommand{\bibinfo}[2]{#2}
\providecommand{\BIBentrySTDinterwordspacing}{\spaceskip=0pt\relax}
\providecommand{\BIBentryALTinterwordstretchfactor}{4}
\providecommand{\BIBentryALTinterwordspacing}{\spaceskip=\fontdimen2\font plus
\BIBentryALTinterwordstretchfactor\fontdimen3\font minus
  \fontdimen4\font\relax}
\providecommand{\BIBforeignlanguage}[2]{{%
\expandafter\ifx\csname l@#1\endcsname\relax
\typeout{** WARNING: IEEEtran.bst: No hyphenation pattern has been}%
\typeout{** loaded for the language `#1'. Using the pattern for}%
\typeout{** the default language instead.}%
\else
\language=\csname l@#1\endcsname
\fi
#2}}
\providecommand{\BIBdecl}{\relax}
\BIBdecl

\bibitem{Haykin2005}
S.~{Haykin}, ``Cognitive radio: brain-empowered wireless communications,''
  \emph{{IEEE} J. Sel. Areas Commun.}, vol.~23, no.~2, pp. 201--220, Feb. 2005.

\bibitem{Yucek2009}
T.~{Y\"{u}cek} and H.~Arslan, ``A survey of spectrum sensing algorithms for
  cognitive radio applications,'' \emph{{IEEE} Commun. Surveys Tuts.}, vol.~11,
  no.~1, pp. 116--130, Mar. 2009.

\bibitem{Tandra2008}
R.~Tandra and A.~Sahai, ``{SNR} walls for signal detection,'' \emph{{IEEE} J.
  Sel. Topics Signal Process.}, vol.~2, no.~1, pp. 4--17, Feb. 2008.

\bibitem{Gardner1987a}
W.~A. {Gardner}, W.~A. {Brown}, and C.-K. {Chen}, ``Spectral correlation of
  modulated signals: Part {II}--digital modulation,'' \emph{{IEEE} Trans.
  Commun.}, vol.~35, no.~6, pp. 595--601, Jun. 1987.

\bibitem{Quan2008}
Z.~Quan, S.~Cui, H.~Poor, and A.~Sayed, ``Collaborative wideband sensing for
  cognitive radios,'' \emph{{IEEE} Signal Process. Mag.}, vol.~25, no.~6, pp.
  60--73, Nov. 2008.

\bibitem{Taherpour2010}
A.~{Taherpour}, M.~{Nasiri-Kenari}, and S.~{Gazor}, ``Multiple antenna spectrum
  sensing in cognitive radios,'' \emph{{IEEE} Trans. Wireless Commun.}, vol.~9,
  no.~2, pp. 814--823, Feb. 2010.

\bibitem{Tugnait2012}
J.~K. Tugnait, ``On multiple antenna spectrum sensing under noise variance
  uncertainty and flat fading,'' \emph{{IEEE} Trans. Signal Process.}, vol.~60,
  no.~4, pp. 1823 --1832, Apr. 2012.

\bibitem{Sadeghi2008}
H.~Sadeghi and P.~Azmi, ``A novel primary user detection method for
  multiple-antenna cognitive radio,'' in \emph{Proc. International Symposium on
  Telecommunications}, Aug. 2008, pp. 188 --192.

\bibitem{Chen2008}
X.~{Chen}, W.~{Xu}, Z.~{He}, and X.~{Tao}, ``Spectral correlation-based
  multi-antenna spectrum sensing technique,'' in \emph{Proc. IEEE WCNC}, Las
  Vegas, NV, USA, Mar. 31--Apr. 3, 2008.

\bibitem{Jitvanichphaibool2010a}
K.~{Jitvanichphaibool}, Y.-C. {Liang}, and Y.~{Zeng}, ``Spectrum sensing using
  multiple antennas for spatially and temporally correlated noise
  environments,'' in \emph{Proc. IEEE DySPAN}, Singapore, Apr. 6--9, 2010.

\bibitem{Schell1990a}
S.~{Schell} and W.~{Gardner}, ``Detection of the number of cyclostationary
  signals in unknown interference and noise,'' in \emph{Proc. ACSSC}, Pacific
  Grove, CA, USA, Nov. 5--7, 1990.

\bibitem{Lawley1959}
D.~N. {Lawley}, ``Tests of significance in canonical analysis,''
  \emph{Biometrika}, vol.~46, no. 1/2, pp. 59--66, Jun. 1959.

\bibitem{Rebeiz2011}
E.~{Rebeiz} and D.~{Cabric}, ``Low complexity feature-based modulation
  classifier and its non-asymptotic analysis,'' in \emph{Proc. IEEE GLOBECOM},
  Houston, TX, USA, Dec. 5--9, 2011.

\bibitem{Zeng2009}
Y.~{Zeng} and Y.-C. {Liang}, ``Eigenvalue-based spectrum sensing algorithms for
  cognitive radio,'' \emph{{IEEE} Trans. Commun.}, vol.~57, no.~6, pp.
  1784--793, Jun. 2009.

\end{thebibliography}
\end{document}